# Formation and annihilation of nanocavities during keV ion irradiation of Ge

J. C. Kim[*]

*Department of Physics and Seitz Materials Research Laboratory, University of Illinois, Urbana, Illinois 61801*

David G. Cahill and R. S. Averback

*Department of Materials Science and Engineering, and Seitz Materials Research Laboratory, University of Illinois, Urbana, Illinois 61801*

[*] Corresponding author. Present address: Department of Physics, Utah State University, UT 84322. Tel: 435-797-7070; FAX: 435-797-2492; E-mail: jkim14@staff.uiuc.edu

**Abstract**

Nanocavities in Ge(111) created by 5 keV Xe ion irradiation are characterized by *ex situ* transmission electron microscopy and Rutherford backscattering spectrometry. Nanocavities nucleate near the surface and then undergo thermal migration. Nanocavities with average diameter of 10 nm and areal density of $5.1 \times 10^{-3}$ nm$^{-2}$ are observed at 500 $^{\mathrm{o}}$C, while nanocavities with average diameter of 2.9 nm and areal density of $3.1 \times 10^{-3}$ nm$^{-2}$ are observed at 400 $^{\mathrm{o}}$C. The estimated Xe gas pressure inside the nanocavities is 0.035 GPa at 500 $^{\mathrm{o}}$C, much smaller than the estimated equilibrium pressure 0.38 GPa. This result suggests that the nanocavities grow beyond equilibrium size at 500 $^{\mathrm{o}}$C. The nanocavities are annihilated at the surface to form surface pits by the interaction of displacement cascades of keV Xe ions with the nanocavities. These pits are characterized by *in situ* scanning tunneling microscopy. Pits are created on Ge(111) and Ge(001) at temperatures ~ 250-305 $^{\mathrm{o}}$C by keV Xe ions even when less than a bilayer (monolayer) of surface material is removed.

PACS numbers: 61.80.Jh; 61.82.Fk; 61.72.Ff; 68.47.Fg

# I. Introduction

Ion sputtering of solids is used in many fields of science and technology: e.g., thin film microanalysis, preparation of clean surfaces for surface science experiments, sputter deposition, and dry etching in microelectronic device fabrications. Ion sputtering often introduces defects in the crystal lattice such as dislocations and cavities.[1] The application of nanometer-size cavities or *nanocavities* for gettering impurities in Si has been recently investigated.[2,3]

Nanocavities are formed during inert gas ion irradiation due to low solubility of inert gases in solids; for example, nanocavities have been observed in Ni,[1,4] Zr and Zr alloys,[5,6] Nb,[6] Al,[1,7] mica,[8] amorphous Ge,[9-11] and Si.[2,3,12-15] Nanocavities migrate inside the crystal during thermal annealing[6,16] and by the interaction of the nanocavities with high energy ions[17-19] or high energy electrons.[20] During high energy ion irradiation, nanocavities can migrate when nanocavities and displacement cascades overlap; Donnelly *et al.* observed discrete jumps in Au of nanocavities containing He under 400 keV Ar ion irradiation at 230 °C.[17,18] Donnelly *et al.* also observed the disappearance of many nanocavities and proposed that some nanocavities annihilate on the surface while most nanocavities are disintegrated by displacement cascades.[17,18]

In present work, we consider the formation of nanocavities during keV Xe ion irradiation of Ge, and examine their formation as a function of temperature and their interactions with surfaces. We have previously reported the surface morphology of Ge(111) during keV Xe ion etching at temperatures from room temperature to 300 °C.[21]

In this study of the surface morphology of Ge(111), we prepared the starting surfaces by annealing the Ge(111) samples at 620 °C to avoid formation of nanocavities, and we could determine the mechanisms of roughening and smoothening of Ge(111) surfaces during keV Xe ion etching.[21] Therefore, the present work on nanocavities as bulk defects in Ge is complementary to our previous study on the surface morphology of Ge(111) which is governed by the kinetics of surface or near-surface defects.

## II. Experimental details

Our Ge samples are typically $1.5 \times 1.5$ cm$^2$ and are bonded with In to a Mo sample stage which is attached to a 3.5 inch diameter Mo sample block. The sample stage is thermally isolated from the sample block by alumina spacers to enhance the heating efficiency by an electron beam. Sample temperatures are measured by an infrared pyrometer operating with a wavelength band centered at ~ 5 μm. The uncertainty in the temperature measurements is ± 5 °C; the reproducibility is ± 2 °C.

The Xe ion beam is produced by an Omicron ISE 10 sputter ion gun, which generates ions with energies up to 5 keV. For the experiments on the formation of nanocavities, which are performed using transmission electron microscopy (TEM) and Rutherford backscattering spectrometry (RBS), the 5 keV Xe ion flux is $1.0 \times 10^{13}$ ions cm$^{-2}$ s$^{-1}$. For the experiments on the annihilation of nanocavities, which are carried out by scanning tunneling microscopy (STM), the 5 keV Xe ion flux is $3.1 \times 10^{13}$ ions cm$^{-2}$ s$^{-1}$. The angle of incidence of keV Xe ions is 50° from the surface normal in both cases.

The Ge(111) and Ge(001) surfaces for STM experiments are prepared by 5 keV Xe ion etching at 520 $^o$C with an ion fluence of $5.6 \times 10^{16}$ ions cm$^{-2}$, corresponding to a removal of 120 bilayers of Ge(111) and a removal of 270 monolayers of Ge(001).[22] This high temperature ion etching of the Ge surfaces results in crystalline starting surfaces without prominent surface defects other than steps, see Fig. 1. Immediately after the high temperature ion etching, the sample is cooled to a lower temperature, and ion-etched or annealed at that lower temperature. Then the sample is cooled to room temperature, transferred to the STM and imaged. All the STM images in this report are filled state images with a typical bias of 1.8-2.0 V.

## III. Results and discussion

### A. Formation of nanocavities in crystalline Ge

Figure 2 shows cross section TEM micrographs of Ge(111) irradiated by 5 keV Xe ions with an ion fluence of $1.8 \times 10^{16}$ ions cm$^{-2}$ at 400 and 500 $^o$C. More and larger nanocavities are formed at 500 $^o$C than at 400 $^o$C. Even though nanocavities are formed below the surface, Ge(111) surfaces do not show noticeable surface defects created by ion irradiation, see Fig. 1. Diameters and average densities of the nanocavities, measured by TEM, are plotted as a function of depth in Fig. 3. The average diameter of the nanocavities is 10 nm at 500 $^o$C and 2.9 nm at 400 $^o$C. The nanocavities closest to the surface are observed ~ 4 nm below the surface, comparable to the predicted penetration depth of 5 keV Xe ions in Ge(111). Nanocavities are also observed as deep as 550 nm from the surface at 500 $^{\circ}$C and 32 nm from the surface at 400 $^{\circ}$C.

The nucleation rate of nanocavities should strongly depend on the densities of Xe atoms and vacancies. Therefore, the nucleation rate of the nanocavities containing Xe in Ge(111) must be the highest near the surface where these densities are the highest. Nanocavities nucleated near the surface apparently migrate deeper into the crystal, while some nanocavities may annihilate at the surface. During migration, nanocavities are likely to grow by absorbing nearby vacancies or coalescence with other nanocavities and thus the diffusion coefficient may not be constant.

To derive the diffusion coefficient of nanocavities, we first assume that nanocavities migrate by adatom diffusion on the interior surfaces. The diffusion coefficient of a nanocavity $D_n$ is

$$D_n = \tfrac{1}{6} f_n d^2, \tag{1}$$

where $f_n$ is the jump frequency of the nanocavity and $d$ is the average jump distance.[23] The number of adatoms participating per unit jump of the nanocavity is $(4\pi R^2)n_o$, where $R$ is the radius of the nanocavity and $n_o$ is the equilibrium density of adatoms on the interior surface of the nanocavity. Then,

$$f_n = f_m (4\pi n_o R^2), \tag{2}$$

where $f_m$ is the jump frequency of adatoms. Also, for a given distance traveled by the nanocavity, a total of $4\pi R^3/3\Omega$ atoms are transferred the same distance in the opposite direction, where $\Omega$ is the atomic volume. Therefore,

$$d = a \cdot \frac{3\Omega}{4\pi R^3}, \tag{3}$$

where $a$ is the atomic lattice constant. The surface diffusion coefficient of adatoms $D_m$ is defined[23]

$$D_m = \frac{f_m a^2}{4}. \tag{4}$$

Therefore, from the above equations, the diffusion coefficient of the nanocavity $D_n$ is

$$D_n = \frac{3n_o \Omega^2}{2\pi R^4} \cdot D_m; \tag{5}$$

therefore, larger nanocavities migrate much more slowly than smaller nanocavities.

If we assume $D_n$ is constant, the diffusion distance $l$ of a nanocavity is

$$l = \sqrt{4D_n t} = \frac{\Omega}{R^2} \cdot \sqrt{\frac{6n_o D_m t}{\pi}}, \tag{6}$$

where $t$ is the diffusion time of the nanocavity. The slope of the two straight guidelines drawn in Fig. 3(a) is $-1/2$ and Eq. (6) seems to be consistent with our observation for nanocavities found deep inside the crystal. From the guidelines, the diffusion distance or depth of nanocavities at 500 $^o$C is approximately 100 times larger than that at 400 $^o$C for a similar nanocavity diameter. This result implies that, from Eq. (6), the product of equilibrium density of adatoms and surface diffusion coefficient at 500 $^o$C is larger than that at 400 $^o$C by a factor of $10^4$.

Here, we estimate $n_o D_m$ assuming the nanocavities migrate by adatom diffusion on the interior surface of nanocavities. In Fig. 3(a), extrapolation of the guideline for the 400 $^o$C data points crosses ($l = 100$ nm, $R = 1$ nm), and therefore for $t = 30$ min, $n_o D_m \approx 6$ $\times 10^3$ s$^{-1}$ at 400 $^o$C. The product of equilibrium density of adatoms and surface diffusion coefficient at 500 $^o$C is $\sim 10^4$ times larger than that at 400 $^o$C; therefore $n_oD_m \approx 6 \times 10^7$ s$^{-1}$ at 500 $^o$C.

Schwarz-Selinger et al.[24] determined the "transport rate" $n_oD_m$ for surface mass transport by ad-dimers on a Ge(001) wetting layer, $n_oD_m = 4.3 \times 10^2$ s$^{-1}$ at 400 $^o$C and 8.6 $\times 10^3$ s$^{-1}$ at 500 $^o$C. Considering possible errors in the measurement of diameters and depth of nanocavities using TEM and the assumption of constant $D_n$, the estimated $n_oD_m$ is probably in reasonable agreement with Schwarz-Selinger et al.'s transport rate at 400 $^o$C. However, the discrepancy of a factor of $10^4$ between our estimate of $n_oD_m$ and Schwarz-Selinger et al.'s transport rate at 500 $^o$C is too large, and our assumption of constant $R$ is probably incorrect. Smaller nanocavities formed near the surface may migrate deep into the crystal and grow by absorbing vacancies or coalescence with other nanocavities; thus, large nanocavities can be observed with apparently greater than predicted diffusion distance.

Figure 4 displays RBS spectra of the Ge(111) irradiated by 5 keV Xe ions at 400, 500 and 600 $^o$C; the ion fluence is $1.8 \times 10^{16}$ ions cm$^{-2}$ and Xe peaks are marked by arrows. The areal density of Xe atoms remaining in the crystal is $1.2 \times 10^{15}$ cm$^{-2}$ at 400 $^o$C, $1.6 \times 10^{15}$ cm$^{-2}$ at 500 $^o$C and less than $10^{14}$ cm$^{-2}$ at 600 $^o$C. In the RBS spectrum, Xe atoms are detected as deep as 32 nm at 400 $^o$C, which agrees well with the depth of nanocavity distribution measured by TEM, see Fig. 3. At 500 $^o$C, due to the high diffusivity of Xe atoms and nanocavities, a broadened Xe peak is observed. At 600 $^o$C, due to the higher diffusivity of Xe atoms and nanocavities, no Xe peak is observed;

however, we found that nanocavities are still formed at 600 $^{o}$C with number density smaller than at 400 and 500 $^{o}$C by an order of magnitude.

To estimate the average density of Xe atoms in the crystal, we measured the size of nanocavities under a selected surface area: the surface area, measured by TEM, is (horizontal length scale) × (thickness of TEM foil), and using the horizontal length scales, we measured the diameters of nanocavities, see Fig. 3(a). The surface area under which we obtained data points for 400 $^{o}$C in Fig. 3(a) is $1.0 \times 10^4$ nm$^2$. If we assume that the nanocavities observed under this surface area are filled with Xe atoms with the density of solid Xe, $2.0 \times 10^{22}$ cm$^{-3}$, then the total number of Xe atoms is $1.4 \times 10^4$, and the areal density of Xe atoms is $1.4 \times 10^{14}$ cm$^{-2}$. We compare the latter with the RBS result of $1.2 \times 10^{15}$ cm$^{-2}$. Therefore, at most 10 % of Xe atoms detected by RBS can be inside the observed nanocavities at 400 $^{o}$C, and small nanocavities containing Xe atoms are apparently present but not detected by TEM.

The average density of Xe atoms for the sample prepared at 500 $^{o}$C is $3.3 \times 10^{21}$ cm$^{-3}$; the total number of Xe atoms is (surface area) × (density of Xe atoms), where the surface area was measured by TEM, $1.1 \times 10^4$ nm$^2$, and the atomic density of Xe atoms was measured by RBS, $1.6 \times 10^{15}$ cm$^{-2}$, and the total volume of the nanocavities is calculated using nanocavity diameters measured by TEM, see Fig. 3(a).

To estimate the Xe gas pressure inside the nanocavities formed at 500 $^{o}$C, we use van der Waals gas approximation. The van der Waals equation of state is

$$P_g = \frac{NkT}{V(1 - \frac{N}{V}B)} \approx \frac{NkT}{V}(1 + \frac{B}{(V/N)}), \tag{7}$$

where $P_g$ is the gas pressure, $N$ is the total number of Xe atoms inside the nanocavity, $V$ is the volume of the nanocavity, and $B$ is a virial coefficient.[25,26] The virial coefficient $B$ for Xe gas at 500 $^o$C is $-0.13$ cm$^3$ mol$^{-1}$,[26] and using the average density of Xe atoms $B/(V/N) = 7.1 \times 10^{-4}$. The estimated van der Waals Xe gas pressure inside the nanocavities at 500 $^o$C is then 0.035 GPa; the estimated ideal Xe gas pressure inside the nanocavities at 500 $^o$C is also 0.035 GPa since $B/(V/N)$ is small.

Equilibrium of a nanocavity is reached when the pressure exerted on the cavity wall by the Xe gas equals the pressure caused by surface tension, i.e., the equilibrium pressure $p$ is

$$p = \frac{2\gamma}{R}, \tag{8}$$

where $\gamma$ is the surface energy. To estimate the average equilibrium pressure, we calculated $<R>$, which is defined as $4\pi<R>^3/3 =$ (total nanocavity volume observed by TEM). The surface energy of Si(111) is 1.2 J/m$^2$;[27] using the ratio of the cohesive energy of Si to Ge, 1.2, we estimate $\gamma \approx 1$ J/m$^2$ for Ge. With $<R> = 5.2$ nm, $p = 0.38$ GPa. The equilibrium pressure is therefore much larger than the estimated van der Waals Xe gas pressure 0.035 GPa, which suggests that nanocavities grow beyond their equilibrium size at 500 $^o$C.

Two mechanisms are possible candidates for the growth of nanocavities beyond the equilibrium size: i) bias driven growth[1,28] and ii) growth due to the coalescence of nanocavities.[29] Here, bias driven growth suggests that, in the presence of biased sinks of point defects, bulk vacancies can be readily absorbed by nanocavities allowing the nanocavities to grow. TEM micrographs in Fig. 2 do not reveal bulk defects such as dislocations that can act as biased sinks of interstitials for bias driven growth.[1,28] Therefore, if bias causes the growth of nanocavities, possible sources for the bias are the Ge surface acting as biased interstitial sink and the nanocavities acting as biased vacancy sinks.

## B. Annihilation of nanocavities on Ge

We discovered that the nanocavities formed at 520 $^o$C can be annihilated at the surface by ion irradiation at lower temperatures, ~ 250-305 $^o$C. Figure 5 shows STM images of the Ge surfaces following 5 keV Xe ion etching for 18 seconds with an ion fluence of $5.6 \times 10^{14}$ ions cm$^{-2}$; this ion fluence corresponds to 1.2 bilayer removal for Ge(111) and 2.7 monolayers removal for Ge(001).[22] Pits are formed on the Ge surfaces while no pits are observed prior to this brief ion exposure, see for example Fig. 1. Small area scans reveal that the pits are surrounded by closely spaced steps. The size of the pits is larger at 305 $^o$C than at 275 $^o$C, and larger on Ge(111) than on Ge(001); the average diameter of the pits on Ge(111) is 24 nm at 275 $^o$C and 40 nm at 305 $^o$C.

We count the number of the pits from STM images and plot the areal density of pits in Fig. 6 as a function of thickness removed for 650 eV, 5 keV and 20 keV Xe ion etching experiments. We observed that the pits completely disappear on Ge(111) after 5 keV Xe ion etching removal of 260 nm-thickness. More pits are formed on Ge(001) than Ge(111). Also, more pits are formed at 275 $^o$C than at 305 $^o$C.

To confirm that the pits are not surface craters as observed by Bellon *et al*. on Ge(001) at room temperature after 20 keV Ga ion irradiation,[30] we studied 5 keV Xe ion etching of Ge without nanocavities below the surface. We grew a Ge(111) buffer layer at 365 $^\circ$C with the thickness of 100 nm in an *in situ* MBE chamber. 5 keV Xe ion etching at 275 $^\circ$C of the Ge(111) buffer layer revealed vacancy islands but no pits, and therefore the pits are not surface craters created by microexplosion[30] or viscous flow[31] through single ion displacement cascades.

The areal density of the pits formed on Ge(111) at 275 $^\circ$C by 5 keV Xe ions and the areal density of the nanocavities formed in Ge(111) at 500 $^\circ$C are plotted as a function of depth or thickness removed in Fig. 7. The similarity between these two plots suggests that the pits are formed by annihilation of the nanocavities at the surface.

To examine whether the pits are formed by thermal migration of the nanocavities, we annealed the Ge(111) starting surface, see Fig. 1(a), for 18 seconds at 305 $^\circ$C; this annealing did not produce pits or vacancy islands. Therefore, pits are not formed by thermal migration of the nanocavities, in agreement with expectations based on the diffusion coefficient of the nanocavities found in part A.

To study the dependence of formation of pits on ion energy, we examined the effects of 650 eV and 20 keV Xe ion irradiation on Ge(111) with nanocavities formed at 520 $^\circ$C. 20 keV Xe ions create about 4 times as many pits as 5 keV Xe ions on Ge(111) during removal of ~ 0.1 nm-thickness, see Fig. 6. Also, the number of the pits created by 20 keV Xe ions starts to decrease at a lower ion fluence than the number of the pits created by 5 keV Xe ions. We believe the difference in the number of pits after removal of ~ 0.1 nm-thickness and earlier decrease in the number of pits during 20 keV Xe ion irradiation are due to the difference in the depth of displacement cascades; since 20 keV Xe ions penetrate deeper into Ge, more nanocavities are annihilated and appear as pits after removal of ~ 0.1 nm-thickness and eventually less pits appear on the surface after removal of > 1 nm-thickness.

We propose that the pits are produced due to migration of the nanocavities toward the surface through the interaction of subsurface displacement cascades with nanocavities. Okuniewski *et al.* have shown, for example, that 2 keV Xe ions directed at nanocavities in amorphous Si can transport atoms toward the nanocavities and fill them in a microexplosion event.[32] Along similar lines, Donnelly *et al.*[17,18] observed that a displacement cascade developing adjacent to a nanocavity may form a melt zone that will allow the nanocavity to deform into the molten region. If this melt zone intersects the surface, the nanocavity can be pulled out toward and annihilated at the surface during recrystallization. Once appearing on the surface, pits flatten due to thermal diffusion of surface vacancies or adatoms, and thus larger pits are formed at 305 $^{\circ}$C than at 275 $^{\circ}$C on both Ge(111) and Ge(001), see Fig. 5.

This proposed formation mechanism of the pits also explains the starting surfaces without pits and initial increase and eventual decrease in the number of the pits. Once the nanocavities appear on the surface at 520 $^{\circ}$C, pits flatten rapidly and therefore no pits are observed on the Ge starting surfaces. The number of the pits initially increases and eventually decreases with increasing ion fluence as the nanocavities are consumed through annihilation at the surface.

## IV. Conclusion

We observed formation of nanocavities containing Xe in crystalline Ge during ion irradiation at temperatures 400-600 $^o$C. Nanocavities nucleate near the surface and then migrate as far as 550 nm from the surface at 500 $^o$C and up to 32 nm from the surface at 400 $^o$C. These nanocavities are annihilated at the surface to appear as pits during subsequent ion etching at lower temperatures ~ 250-305 $^o$C due to the interaction between displacement cascades and nanocavities.

## Acknowledgements

This work was supported by the U.S. National Science Foundation Grant No. DMR-9986160. TEM and RBS experiments were carried out in the Center for Microanalysis of Materials, University of Illinois, which is supported by the U.S. Department of Energy under grant DEFG02-91-ER45439.

**References**

[1] D. I. R. Norris, Rad. Effects **14**, 1 (1972).

[2] E. Chason, S. T. Picraux, J. M. Poate, J. O. Borland, M. I. Current, T. D. de la Rubia, D. J. Eaglesham, O. W. Holland, M. E. Law, C. W. Magee, J. W. Mayer, J. Melngailis, and A. F. Tasch, J. Appl. Phys. **81**, 6513 (1997).

[3] G. A. Petersen, S. M. Myers and D. M. Follstaedt, Nucl. Instr. Meth. Phys. Res. B **127/128**, 301 (1997).

[4] J. Rest and R. C. Birtcher, J. Nucl. Mat. **168**, 312 (1989).

[5] L. Pagano Jr., A. T. Motta, and R. C. Birtcher, J. Nucl. Mater. **244**, 295 (1997).

[6] P. J. Goodhew and S. K. Tyler, Proc. R. Soc. Lond. A **377**, 151 (1981).

[7] N. Ishikawa, M. Awaji, K. Furuya, R. C. Birtcher, and C. W. Allen, Nucl. Instr. Meth. Phys. Res. B **127/128**, 123 (1997).

[8] G. A. Hishmeh, L. Cartz, F. Desage, C. Templier, J. C. Desoyer, and R. C. Birtcher, J. Mater. Res. **9**, 3095 (1994).

[9] L. M. Wang and R. C. Birtcher, Appl. Phys. Lett. **55**, 2494 (1989).

[10] L. M. Wang and R. C. Birtcher, Philos. Mag. A **64**, 1209 (1991).

[11] H. Huber, W. Assmann, S. A. Karamian, A. Mücklich, W. Prusseit, E. Gazis, R. Grötzchel, M. Kokkoris, E. Kossiondis, H. D. Mieskes, R. Vlastou, Nucl. Instr. Meth. Phys. Res. B **122**, 542 (1997).

[12] V. Raineri, S. U. Campisano, Nucl. Instr. Meth. Phys. Res. B **120**, 56 (1996).

[13] V. Raineri, S. U. Campisano, Appl. Phys. Lett. **69**, 1783 (1996).

[14] V. Raineri and M. Saggio, Appl. Phys. Lett. **71**, 1673 (1997).

[15] C. C. Griffioen, J. H. Evans, P. C. De Jong and A. Van Veen, Nucl. Instr. Meth. Phys. Res. B **27**, 417 (1987).

[16] H. Yamaguchi, I. Hashimoto, H. Mitsuya, K. Nakamura, E. Yagi and M. Iwaki, J. Nucl. Mater. **161**, 164 (1989).

[17] S. E. Donnelly, R. C. Birtcher, C. Templier, and V. Vishnyakov, Phys. Rev. B **52**, 3970 (1995).

[18] S. E. Donnelly, R. C. Birtcher, C. Templier, R. Valizadeh, and V. Vishnyakov, Mat. Res. Soc. Symp. Proc. **373**, 243 (1995).

[19] D. E. Alexander and R. C. Birtcher, J. Nucl. Mater. **191-194**, 1289 (1992).

[20] C. W. Allen, R. C. Birtcher, S. E. Donnelly, K. Furuya, N. Ishikawa, and M. Song, Appl. Phys. Lett. **74**, 2611 (1999).

[21] J. Kim, D. G. Cahill, and R. S. Averback, Phys. Rev. B **67**, 045404 (2003).

[22] A. L. Southern, W. R. Willis, and M. T. Robinson, J. Appl. Phys. **34**, 153 (1963); the sputtering yield of 5 keV Ar ions on Ge(111) is 3 and that on Ge(001) is 3.4. The sputtering yield of Xe ions on Ge and Ar ions on Ge are expected nearly the same since the energy transfer coefficient for both cases are nearly identical. We also used NIH Image program to measure the area of the vacancy islands on Ge(111) produced by 5 keV Xe ion sub-bilayer etching and obtained sputtering yield of 2.5 in reasonable agreement with the above literature value. Therefore, we used sputtering yield of 3 for 5 keV Xe ion etching on Ge in this report.

[23] F. A. Nichols, J. Nucl. Mater. **30**, 143 (1969).

[24] T. Schwarz-Selinger, Y. L. Foo, D. G. Cahill and J. E. Greene, Phys. Rev. B **65**, 125317 (2002).

[25] W. A. Coghlan and L. K. Mansur, J. Nucl. Mat. **122&123**, 495 (1984).

[26] J. H. Dymond and E. B. Smith, *The Virial Coefficients of Pure Gases and Mixtures* (Clarendon Press, Oxford, 1980).

[27] W. Mönch, *Semiconductor Surfaces and Interfaces* (Springer, New York, 2001).

[28] A. Hishinuma and L. K. Mansur, J. Nucl. Mat. **118**, 91 (1983).

[29] D. Preininger and D. Kaletta, J. Nucl. Mat. **117**, 239 (1983).

[30] P. Bellon, S. Jay Chey, J. E. Van Nostrand, M. Ghaly, D. G. Cahill, and R. S. Averback, Surf. Sci. **339**, 135 (1995).

[31] C. Teichert, M. Hohage, T. Michely, and G. Comsa, Phys. Rev. Lett. **72**, 1682 (1994).

[32] M. A. Okuniewski, Y. Ashkenazy, B. Heuser, and R. S. Averback, unpublished.

**Figure captions**

Fig. 1.  STM images of (a) Ge(111) and (b) Ge(001) surfaces etched by 5 keV Xe ions for 30 min at 520 $^o$C with the ion fluence of $5.6 \times 10^{16}$ ions cm$^{-2}$. The scan size is $600 \times 600$ nm$^2$.

Fig. 2.  Cross sectional TEM micrographs of Ge(111) irradiated by 5 keV Xe ions with an ion fluence of $1.8 \times 10^{16}$ ions cm$^{-2}$ at (a) 400 $^o$C and (b) 500 $^o$C. The Ge(111) surfaces are marked by arrows.

Fig. 3.  (a) Diameters of the nanocavities observed in Fig. 2 as a function of depth. 56 nanocavities with average diameter of 10 nm are observed at 500 $^o$C below the surface area of $1.1 \times 10^4$ nm$^2$; 31 nanocavities with average diameter of 2.9 nm are observed at 400 $^o$C below the surface area of $1.0 \times 10^4$ nm$^2$. (b) Average density of the nanocavities observed in Fig. 2 as a function of depth.

Fig. 4.  RBS data for Ge(111) irradiated by 5 keV Xe ions at 400, 500 and 600 $^o$C with an ion fluence of $1.8 \times 10^{16}$ ions cm$^{-2}$. Ge surface and Xe peaks are marked by arrows.

Fig. 5.  STM images of the Ge surfaces etched by 5 keV Xe ions for 18 s with an ion flux of $3.1 \times 10^{13}$ ions cm$^{-2}$ s$^{-1}$, corresponding to 1.2 bilayer removal from the Ge(111) surfaces in (a) and (b), and 2.7 monolayers removal from the Ge(001) surfaces in (c) and (d). The scan size is $600 \times 600$ nm$^2$. Temperature during ion etching is: (a) 275 $^o$C (b) 305 $^o$C (c) 275 $^o$C (d) 305 $^o$C.

Fig. 6. The number of the pits as a function of the thickness removed. Filled symbols are Ge samples etched by 5 keV Xe ions. Sample orientation and temperature are ▲: (111), 305 $^o$C; ▼: (111), 275 $^o$C; ■: (001), 305 $^o$C; ◆: (001), 275 $^o$C. Open symbols are Ge(111) samples etched by 20 keV Xe ions. Sample temperature is ○: 275 $^o$C, □: 215 $^o$C, △: 245 $^o$C and ▽: 305 $^o$C. × and + are Ge(001) and Ge(111) samples etched by 650 eV Xe ions at 245 $^o$C, respectively. The error bars are calculated by assuming that the statistics of the pits follows a Poisson distribution.

Fig. 7. The areal density of the pits and nanocavities as a function of depth. The areal density of the pits is obtained from STM images as in Fig. 6 and the areal density of the nanocavities is obtained from TEM images. The areal density of the nanocavities is calculated by integrating the average nanocavity density shown in Fig. 3(b) over a 10 nm depth.

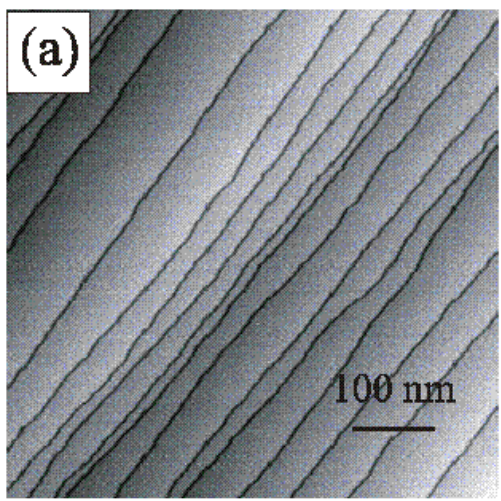

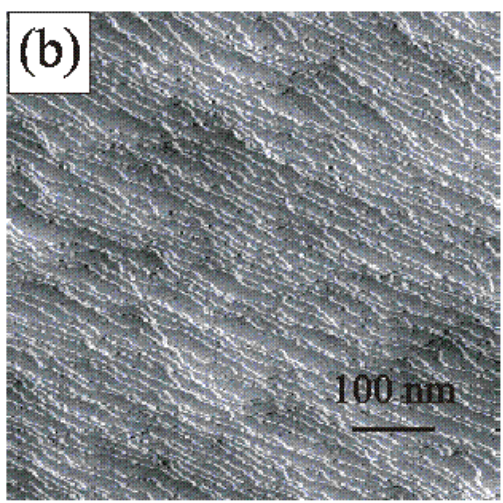

Fig. 1   J. C. Kim, PRB

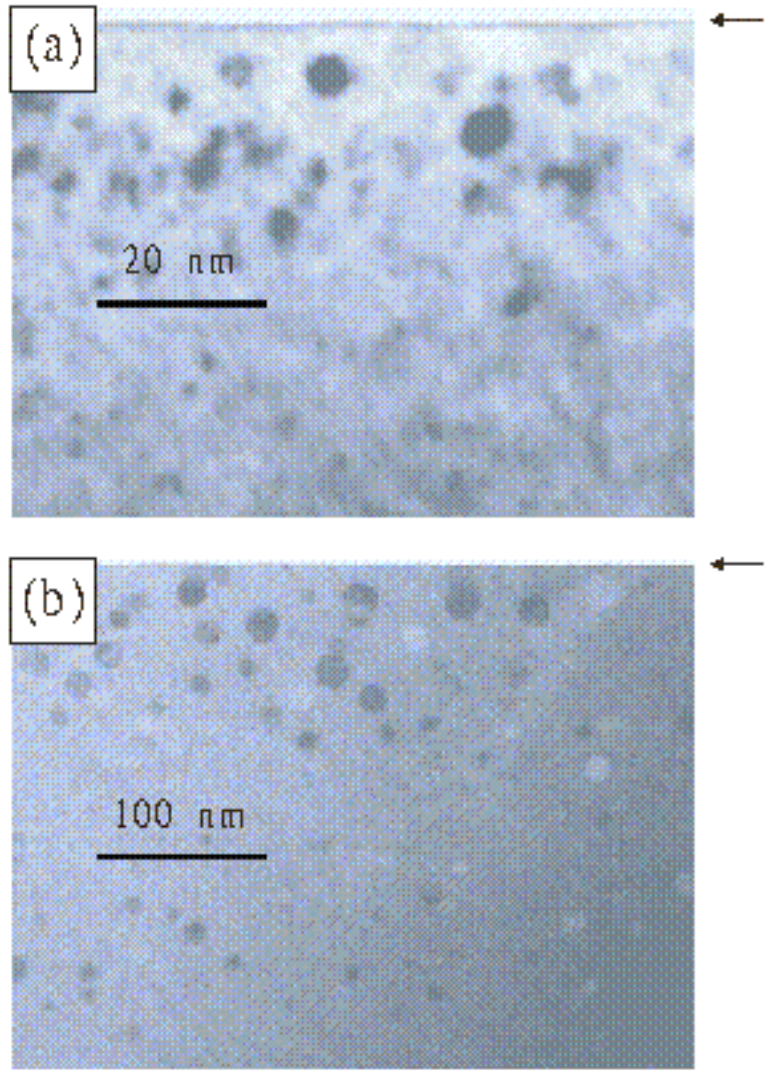

Fig. 2   J. C. Kim, PRB

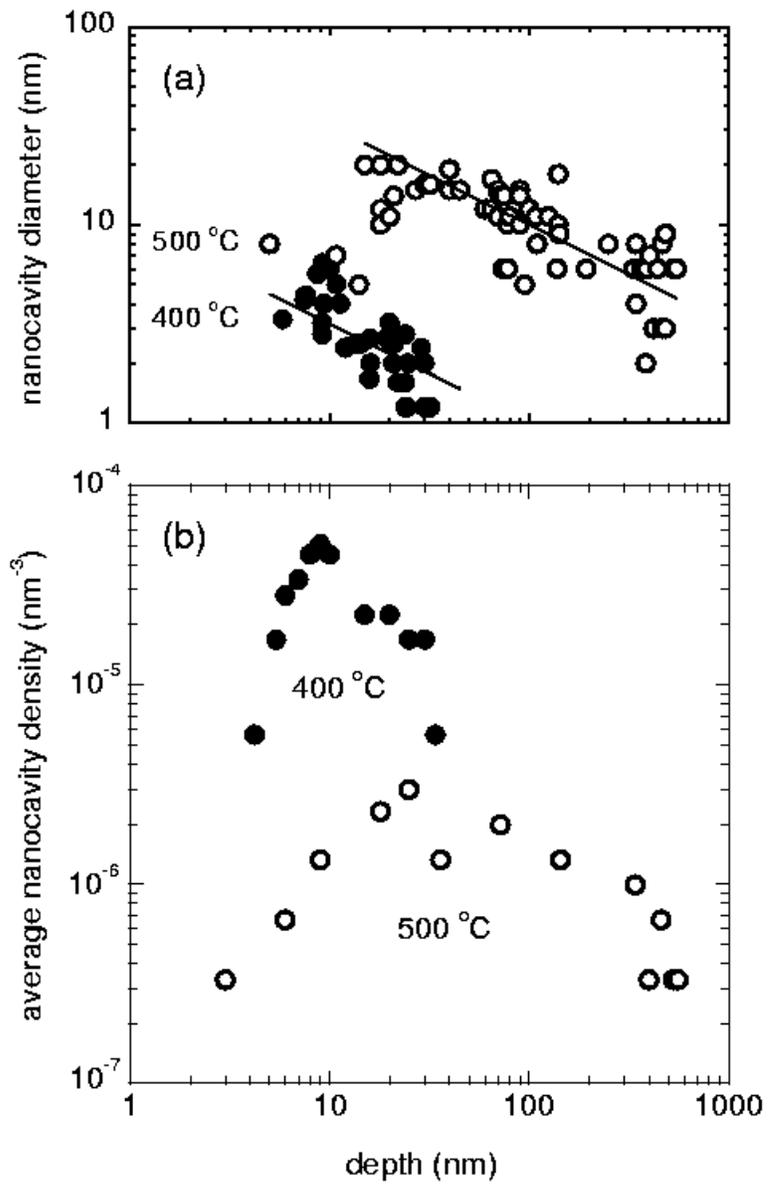

Fig. 3   J. C. Kim, PRB

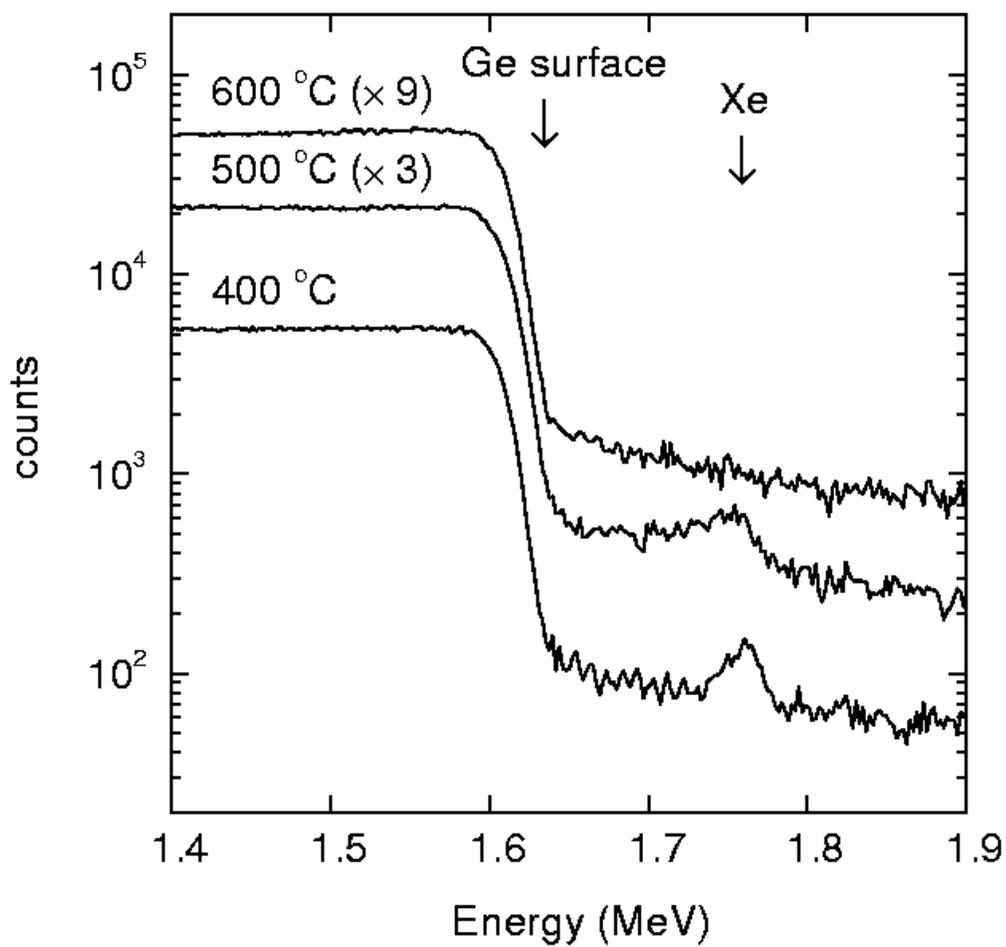

Fig. 4   J. C. Kim, PRB

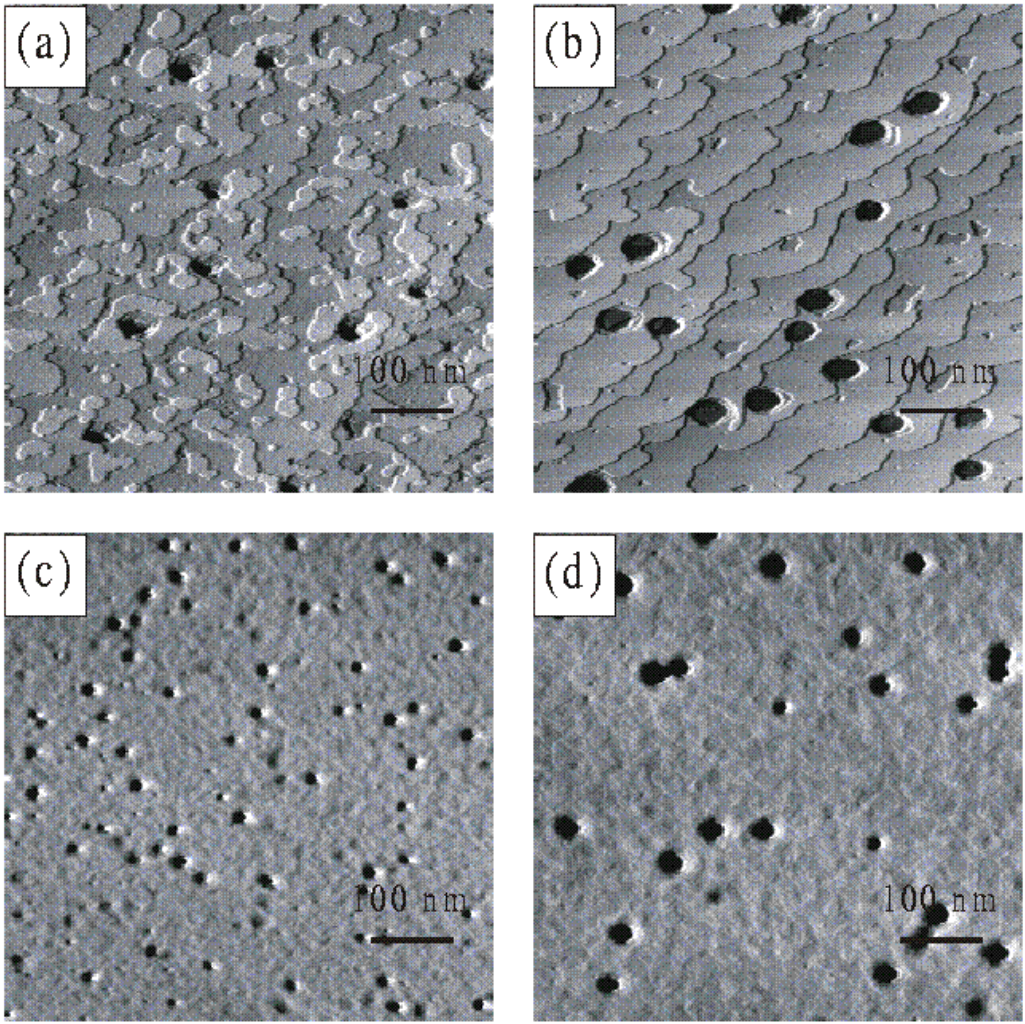

Fig. 5   J. C. Kim, PRB

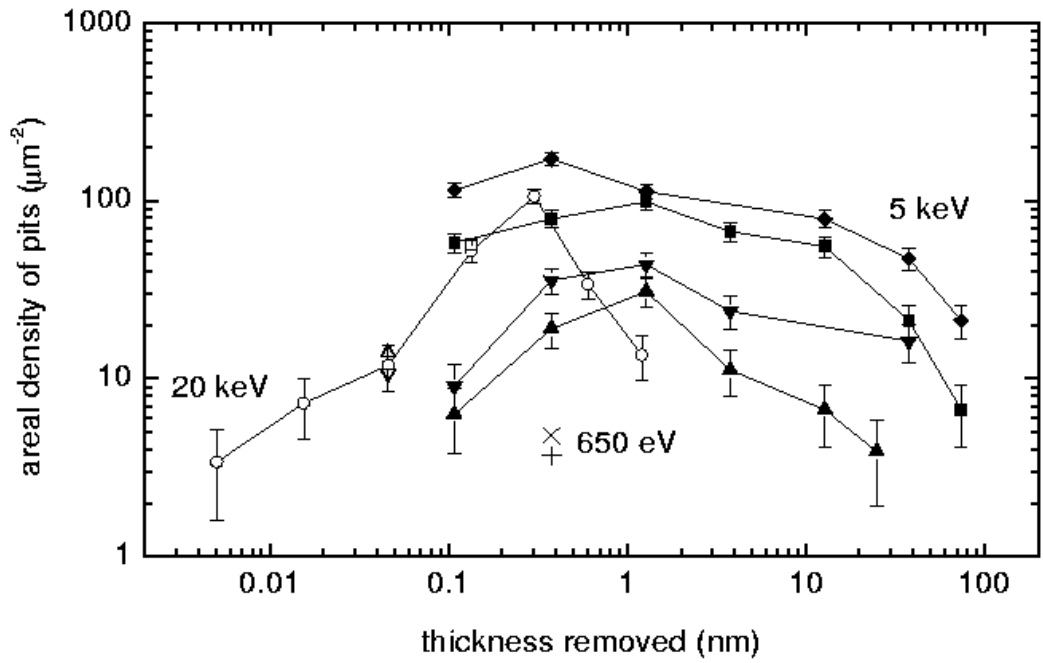

Fig. 6   J. C. Kim, PRB

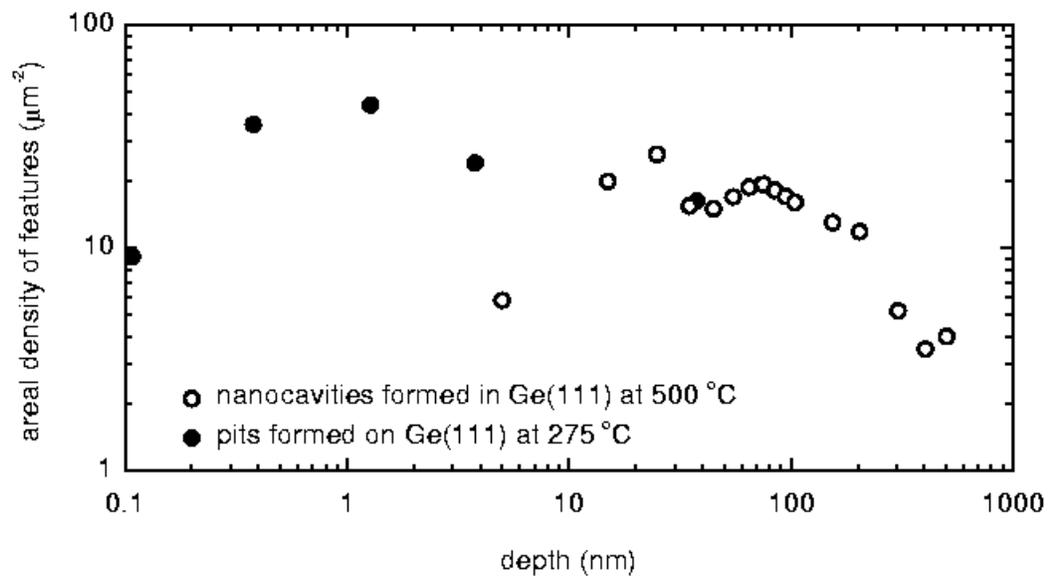

Fig. 7   J. C. Kim, PRB